\documentclass[twocolumn,superscriptaddress,showpacs,preprintnumbers]{revtex4}
\usepackage{amssymb}
\usepackage{amsmath}
\usepackage{graphicx}
\usepackage{dcolumn}
\usepackage{verbatim}
\usepackage{bm}

\begin{document}

\title{Quantum Critical Dynamics of A Qubit Coupled to An Isotropic
Lipkin-Meshkov-Glick Bath}

\author{H.T. Quan}
\affiliation{Department of Physics and Center of Theoretical and
Computational Physics, The University of Hong Kong, Pokfulam Road,
Hong Kong, China} \affiliation{Institute of Theoretical Physics,
Chinese Academy of Sciences, Beijing, 100080, China}
\author{Z.D. Wang}
\affiliation{Department of Physics and Center of Theoretical and
Computational Physics, The University of Hong Kong, Pokfulam Road,
Hong Kong, China}
\author{C. P. Sun}
\affiliation{Institute of Theoretical Physics, Chinese Academy of
Sciences, Beijing, 100080, China}

\begin{abstract}
We explore a dynamic signature of quantum phase transition (QPT)
in an isotropic Lipkin-Meshkov-Glick (LMG) model by studying the
time evolution of a central qubit coupled to it. We evaluate
exactly the time-dependent purity, which can be used to measure
quantum coherence, of the central qubit. It is found that
distinctly different behaviors of the purity as a function of the
parameter reveal clearly the QPT point in the system. It is also
clarified that the present model is equivalent to an anti
Jaynes-Cummings model under certain conditions.
\end{abstract}

\maketitle

\section{INTRODUCTION}

Quantum phase transitions (QPTs) \cite{QPT} in spin systems, e.g., the XY
model \cite{xy}, the Lipkin-Meshkov-Glick (LMG) model \cite{LMG}, and the
Dicke model \cite{dicke}, have aroused much interest in recent years. Most
of these efforts have addressed possible connections of quantum entanglement
measures, such as the concurrence, the entanglement entropy, and the
negativity, with the QPTs in the systems. The scaling behavior \cite%
{xy,scaling behavior} of the entanglement demonstrate well the
quantum criticality of these systems. On the other hand, being
related to quantum measurement theory and quantum decoherence
problems, theoretical studies of the bath influence on the dynamic
property of a central system have also attracted much attention.
Ref. \cite{Wezel} claimed to find \textquotedblleft an intrinsic
limit to quantum coherence due to spontaneous symmetry breaking"
in Lieb-Mattis model; the relationship between entanglement
dynamics and paramagnet-ferromagnet phase transition was explored
in Ref.\ \cite{Paganelli}; it was shown in Ref. \cite{Quan} that
the Loschmidt echo decay enhanced at the critical point can be a
signature of QPT in the transverse Ising model. In fact, further
deeper studies on the the relevant issues of open quantum systems
not only provides us a better understanding of the
quantum-classical crossover, but also promises important potential
applications in quantum information processing \cite{Quantum
information}.

In this paper, integrating coherently the above two interesting topics:
quantum phase transition and quantum open system \cite{Combination}, we
elaborate how the QPT of the \textquotedblleft bath" influences the dynamics
of a central qubit coupled to it. It is shown that when the bath is in
different phases, the purity of the central qubit exhibit distinctly
different behaviors in two different phases. Moreover, it is also
illustrated that under certain conditions our model is equivalent to an anti
Jaynes-Cummings (anti J-C) model \cite{J-C model}.

The paper is organized as follows. In Sec. II, we introduce the LMG model
and summarize the main properties of this model. In Sec. III, we study the
dynamic evolution of a central qubit coupled to a bath described by an
isotropic LMG model, which is exactly solvable. In Sec. IV, We evaluate the
purity of the central qubit in the symmetry broken phase and the symmetric
phase, respectively. The QPT of the bath is well indicated by the behavior
of the purity. In Sec. V, a connection between the current model and an anti
J-C model is established. Section VI presents our summary and conclusion.

\section{LIPKIN-MESHKOV-GLICK (LMG) MODEL FOR QUANTUM PHASE TRANSITION}

We consider a central qubit (two-level system) that couples to a multi-qubit
bath, which is described by the LMG model \cite{LMG}%
\begin{equation}
H_{B}=-\frac{\lambda }{N}\sum_{i<j}^{N}(\sigma _{i}^{x}\sigma
_{j}^{x}+\gamma \sigma _{i}^{y}\sigma _{j}^{y})-\sum_{i=1}^{N}\sigma
_{i}^{z},  \label{1}
\end{equation}%
where $\sigma _{i}^{\alpha }$, $\alpha =x,y,z$ ($i=1,2,\cdots N$) are the
Pauli matrices of the $i$-th atom, $\lambda /N$ denotes the coupling
strength, which is inversely proportional to the atom number $N$. This
Hamiltonian contains long-range interactions, i.e., every spin in the bath
interacts with all the others. In the isotropic case, $\gamma =1$, the
Hamiltonian is diagonal in the Dicke representation%
\begin{equation}
H_{B}=-\frac{2\lambda }{N}\left[ \mathbf{J}_{N}^{2}-(J_{N}^{z})^{2}-\frac{N}{%
2}\right] -2J_{N}^{z},  \label{2}
\end{equation}%
and the ground state of $H_{B}$ lies in the subspace spanned by the Dicke
states $\{\left\vert N/2,M\right\rangle ,M=-N/2,\cdots N/2\}$ \cite{Vidal}.
Here, $\mathbf{s=}\vec{\sigma}/2$, $J_{N}^{\alpha }=1/2\sum_{i=1}^{N}\sigma
_{i}^{\alpha }$ and
\begin{eqnarray}
\mathbf{J}_{N}^{2}\left\vert \frac{N}{2},M\right\rangle &=&\frac{N}{2}\left(
\frac{N}{2}+1\right) \left\vert \frac{N}{2},M\right\rangle ,  \label{3} \\
J_{N}^{z}\left\vert \frac{N}{2},M\right\rangle &=&M\left\vert \frac{N}{2}%
,M\right\rangle .  \notag
\end{eqnarray}%
The eigenenergy corresponding to $\left\vert N/2,M\right\rangle $ is $%
2\lambda M^{2}/N-2M-\lambda N/2$. Hence, the ground state $\left\vert
G\right\rangle $ is $\lambda $-dependent \cite{Dusuel}, i.e.,
\begin{equation}
\left\vert G\right\rangle =\left\{
\begin{array}{c}
\left\vert \frac{N}{2},\frac{N}{2}\right\rangle \,,\text{ }(0<\lambda <1),
\\
\left\vert \frac{N}{2},I(\lambda )\right\rangle ,\text{ }(\lambda >1)%
\end{array}%
\right.  \label{4}
\end{equation}%
where $I(\lambda )$ is the integer nearest to $N/2\lambda $. The level
crossing at $\lambda =1$ leads to the occurrence of a QPT. This point is
also a symmetry breaking point \cite{Dusuel,Symmetry breaking}: when $%
0<\lambda <1 $, the ground state of the bath is unique and fully polarized
in the magnetic field direction, and thus the bath is in a symmetry broken
phase; when $\lambda >1$, the ground state is infinitely degenerate and thus
the bath is in a symmetric phase. We below elaborate how the dynamic
evolution of the purity \cite{Purity} (a measure of quantum coherence)
depends on the coupling strength between the central spin and the bath; in
particular, we observe that the purity shows distinctly different behaviors
in the two phases, which may be used to reveal the QPT point in the bath.

\section{DYNAMICS OF A CENTRAL QUBIT COUPLED TO AN ISOTROPIC LMG MODEL}

A spin-bath model is described by the total Hamiltonian $%
H=H_{B}+H_{S}+H_{SB} $ \cite{Square root,Spin bath}, where $H_{S}=-2s_{z}$
is the free Hamiltonian of the central qubit $S$. $H_{SB}$ denotes the
coupling between $S$ and the bath $B$. Specifically, the total Hamiltonian
can be written as%
\begin{eqnarray}
H &=&-\frac{\lambda }{N}\sum_{i<j}^{N}(\sigma _{i}^{x}\sigma _{j}^{x}+\sigma
_{i}^{y}\sigma _{j}^{y})-\sum_{i=1}^{N}\sigma _{i}^{z}  \label{5} \\
&&+\lambda ^{\prime }\sum_{i}^{N}(\sigma _{i}^{x}\sigma ^{x}+\sigma
_{i}^{y}\sigma ^{y})-\sigma ^{z},  \notag
\end{eqnarray}%
where $\sigma ^{\alpha }$, $\alpha =x,y,z$, are the Pauli operators of the
central qubit; $\lambda ^{\prime }$ is the coupling strength between the
central qubit and the bath. In the Dicke representation, the above
Hamiltonian can be rewritten as%
\begin{eqnarray}
H &=&-\frac{\lambda }{N}\left[ J_{N}^{+}J_{N}^{-}+J_{N}^{-}J_{N}^{+}-N\right]
-2J_{N}^{z}  \label{6} \\
&&-2\lambda ^{\prime }(s_{+}J_{N}^{-}+s_{-}J_{N}^{+})-(2s_{z}),  \notag
\end{eqnarray}%
where $J_{N}^{\pm }=J_{N}^{x}\pm iJ_{N}^{y}$ and $s_{\pm }=s^{x}\pm is^{y}$
are the ladder operators of the $N$-qubit bath and the central qubit,
respectively. For simplicity, we denote the two eigenstate of the central
qubit as $\left\vert \uparrow \right\rangle =\left\vert 1/2,1/2\right\rangle
$ and $\left\vert \downarrow \right\rangle =\left\vert 1/2,-1/2\right\rangle
$, and $s_{z}\left\vert \uparrow \right\rangle =\left\vert \uparrow
\right\rangle /2$, $s_{z}\left\vert \downarrow \right\rangle =-\left\vert
\downarrow \right\rangle /2$. In an invariant subspace $\mathcal{H}_{%
\mathcal{M}}$ of $H$\ spanned by the ordered basis vector $\{\left\vert
N/2,M\right\rangle \otimes \left\vert \uparrow \right\rangle ,\left\vert
N/2,M+1\right\rangle \otimes \left\vert \downarrow \right\rangle \}$, the
total Hamiltonian can be expressed as a quasidiagonal matrix with the
diagonal blocks%
\begin{equation}
H_{M}=\left[
\begin{array}{cc}
\alpha , & \zeta \\
\zeta , & \beta%
\end{array}%
\right] ,  \label{8}
\end{equation}%
where
\begin{eqnarray}
\alpha &=&-\frac{\lambda }{2N}\left[ N^{2}-4M^{2}\right] -2M-1,  \label{9} \\
\beta &=&-\frac{\lambda }{2N}\left[ N^{2}-4(M+1)^{2}\right] -2(M+1)+1,
\notag \\
\zeta &=&-\lambda ^{\prime }\sqrt{N\left( N+2\right) -4M(M+1)}.  \notag
\end{eqnarray}%
A straightforward calculation determines the two eigenvalues $x_{1}$ and $%
x_{2}$ of $H_{M}$ as
\begin{eqnarray}
x_{1} &=&\frac{1}{2}[(\alpha +\beta )+\sqrt{(\alpha -\beta )^{2}+4\zeta ^{2}}%
],  \label{11} \\
x_{2} &=&\frac{1}{2}[(\alpha +\beta )-\sqrt{(\alpha -\beta )^{2}+4\zeta ^{2}}%
],  \notag
\end{eqnarray}%
and the eigenstate $\left\vert \Psi _{1}\right\rangle $ corresponding to $%
x_{1}$ is
\begin{equation*}
\left\vert \Psi _{1}\right\rangle =a\left\vert \frac{N}{2},M\right\rangle
\otimes \left\vert \uparrow \right\rangle +b\left\vert \frac{N}{2}%
,M+1\right\rangle \otimes \left\vert \downarrow \right\rangle ,
\end{equation*}%
where%
\begin{eqnarray}
a &=&\frac{\zeta }{\sqrt{(\alpha -x_{1})^{2}+\zeta ^{2}}},  \label{12} \\
b &=&\frac{x_{1}-\alpha }{\sqrt{(\alpha -x_{1})^{2}+\zeta ^{2}}}.  \notag
\end{eqnarray}%
We would like to point out that, in the symmetric phase, all $x_{1}$, $x_{2}$%
, $a$, and $b$ are functions of $I(\lambda )$, i.e., $F=F[I\left( \lambda
\right) ]$ with $F=x_{1}$, $x_{2}$, $a$, and $b$. The dynamic evolution
operator $U(t)=\exp [-iHt]$ in the subspace $\mathcal{H}_{\mathcal{M}}$ can
be expressed in terms of $x_{1}$, $x_{2}$, $a$, and $b$%
\begin{equation}
U_{M}(t)=\left[
\begin{array}{cc}
a^{2}e^{-ix_{1}t}+b^{2}e^{-ix_{2}t}, & ab(e^{-ix_{1}t}-e^{-ix_{2}t}) \\
ab(e^{-ix_{1}t}-e^{-ix_{2}t}), & b^{2}e^{-ix_{1}t}+a^{2}e^{-ix_{2}t}%
\end{array}%
\right] ,  \label{10}
\end{equation}

We wish to mention that the above dynamic evolution (\ref{10}) is valid only
for the cases when $-N/2\leq M<N/2$, because $\{\left\vert
N/2,M\right\rangle \otimes \left\vert \uparrow \right\rangle ,\left\vert
N/2,M+1\right\rangle \otimes \left\vert \downarrow \right\rangle \}$ is a
two-dimensional invariant subspace for these cases. But for the case $M=N/2$%
, $\{\left\vert N/2,M\right\rangle \otimes \left\vert \uparrow \right\rangle
\}$ is a one-dimensional invariant subspace, i.e., $\left\vert
N/2,N/2\right\rangle \otimes \left\vert \uparrow \right\rangle $ is an
eigenstate of the total Hamiltonian (\ref{6}), and its corresponding
eigenenergy is $-(N+1)$. Thus the dynamic evolution of this state $%
\left\vert N/2,N/2\right\rangle \otimes \left\vert \uparrow \right\rangle $
is different from Eq. (\ref{10}). We will discuss this point in the next
section.

\section{PURITY OF THE CENTRAL QUBIT AS A WITNESS OF QUANTUM PHASE TRANSITION%
}

Based on the above results, we now solve the Schrodinger equation that
describes the dynamics of the purity of the central qubit. To highlight the
influence of the QPT of the bath on the coupled central qubit, it is assumed
that the bath and the central qubit are initially in the ground state $%
\left\vert G\right\rangle $ (\ref{4}) and a pure superposition state $%
c_{\uparrow }\left\vert \uparrow \right\rangle +c_{\downarrow }\left\vert
\downarrow \right\rangle $, respectively. The evolution of the total system
(the bath plus the central qubit) is%
\begin{equation}
\left\vert \Psi _{N+1}(t)\right\rangle =e^{-iHt}\left\vert G\right\rangle
\otimes (c_{\uparrow }\left\vert \uparrow \right\rangle +c_{\downarrow
}\left\vert \downarrow \right\rangle ),  \label{13}
\end{equation}%
and the reduced density matrix of the central qubit is%
\begin{equation}
\rho ^{S}(t)=\mathrm{Tr_{B}}\left\vert \Psi _{N+1}(t)\right\rangle
\left\langle \Psi _{N+1}(t)\right\vert ,  \label{14}
\end{equation}%
where $\mathrm{Tr_{B}}$ means tracing out the degree of freedom of the bath.

The purity $P$ of the central qubit is defined as%
\begin{eqnarray}
P &=&\mathrm{Tr_{S}}\{[\rho ^{S}(t)]^{2}\}  \label{15} \\
&=&[\rho _{\uparrow \uparrow }^{S}(t)]^{2}+[\rho _{\downarrow \downarrow
}^{S}(t)]^{2}+2\left\vert \rho _{\uparrow \downarrow }^{S}(t)\right\vert
^{2},  \notag
\end{eqnarray}%
which can be used to measure the quantum coherence. For a pure state, the
purity equals to unity, while for a mixed state the purity is less than
unity. The decay of purity indicates the loss of quantum coherence \cite%
{Purity}.

\subsection{Purity of the central qubit in two phases}

\subsubsection{Symmetric phase}

When $\lambda >1$, the bath is in the symmetric phase and $I\left( \lambda
\right) <N/2$. As mentioned above, we can apply the evolution matrix (\ref%
{10}) to obtain the reduced density matrix $\rho ^{S}(t)$ of the central
qubit with the matrix elements defined by%
\begin{eqnarray}
\rho _{\uparrow \uparrow }^{S}(t) &=&\sqrt{[\left\vert c_{\uparrow
}\right\vert ^{2}f(\lambda ,t)+\left\vert c_{\downarrow }\right\vert
^{2}h(\lambda ,t)]^{2}},  \label{16} \\
\rho _{\downarrow \downarrow }^{S}(t) &=&\sqrt{[\left\vert c_{\downarrow
}\right\vert ^{2}g(\lambda ,t)+\left\vert c_{\uparrow }\right\vert
^{2}i(\lambda ,t)]^{2}},  \notag \\
\left\vert \rho _{\uparrow \downarrow }^{S}(t)\right\vert &=&\sqrt{%
\left\vert c_{\uparrow }c_{\downarrow }^{\ast }\right\vert ^{2}g(\lambda
,t)\times f(\lambda ,t)}=\left\vert \rho _{\downarrow \uparrow
}^{S}(t)\right\vert ,  \notag
\end{eqnarray}%
where%
\begin{eqnarray}
f(\lambda ,t) &=&a^{4}+b^{4}+2a^{2}b^{2}\cos [(x_{1}-x_{2})t],  \label{16.3}
\\
g(\lambda ,t) &=&(a^{\prime })^{4}+(b^{\prime })^{4}+2(a^{\prime
})^{2}(b^{\prime })^{2}\cos [(x_{1}^{\prime }-x_{2}^{\prime })t],  \notag \\
h(\lambda ,t) &=&2(a^{\prime })^{2}(b^{\prime })^{2}\{1-\cos [(x_{1}^{\prime
}-x_{2}^{\prime })t]\},  \notag \\
i(\lambda ,t) &=&2a^{2}b^{2}\{1-\cos [(x_{1}-x_{2})t]\},  \notag
\end{eqnarray}%
and the parameters $x_{1}^{\prime }$, $x_{2}^{\prime }$, $a^{\prime }$, and $%
b^{\prime }$ are defined by $F^{\prime }=F[I\left( \lambda \right) -1]$ with
$F^{\prime }=x_{1}^{\prime }$, $x_{2}^{\prime }$, $a^{\prime }$, and $%
b^{\prime }$. This subtle change from $(x_{1},x_{2},a,b)$ to $(x_{1}^{\prime
},x_{2}^{\prime },a^{\prime },b^{\prime })$ is due to the fact that $%
\left\vert G\right\rangle \otimes \left\vert \uparrow \right\rangle $ and $%
\left\vert G\right\rangle \otimes \left\vert \downarrow \right\rangle $
belong to two different invariant subspace $\mathcal{H}_{\mathcal{M}}$ and $%
\mathcal{H}_{\mathcal{M-}1}$. If, for simplicity, we assume that the central
qubit is initially in the superposition state $(\left\vert \uparrow
\right\rangle +\left\vert \downarrow \right\rangle )/\sqrt{2}$, we obtain
from Eq. (\ref{15}) the exact expression of the purity of the central qubit%
\begin{eqnarray}
P &=&\frac{1}{4}[f(\lambda ,t)+h(\lambda ,t)]^{2}+\frac{1}{4}[i(\lambda
,t)+g(\lambda ,t)]^{2}  \label{17} \\
&&+\frac{1}{2}g(\lambda ,t)f(\lambda ,t).  \notag
\end{eqnarray}

\subsubsection{Symmetry broken phase}

When $0<\lambda <1$, the bath is in the symmetry broken phase and the ground
state is the fully polarized state $\left\vert G\right\rangle =\left\vert
N/2,N/2\right\rangle $. $\left\vert G\right\rangle \otimes \left\vert
\uparrow \right\rangle $ is an eigenstate of the total Hamiltonian (\ref{5}%
), and its corresponding eigenenergy is $-(N+1)$. Thus the dynamic evolution
of this state is $\exp [i(N+1)t]$. While the dynamic evolution of the other
state $\left\vert G\right\rangle \otimes \left\vert \downarrow \right\rangle
$ can be obtained following the way mentioned above Eq.(\ref{10}) with $%
M=N/2-1$. After a similar procedure, if the central qubit is initially
prepared in the superposition state $(\left\vert \uparrow \right\rangle
+\left\vert \downarrow \right\rangle )/\sqrt{2}$, the exact expression of
the purity in the symmetry broken phase can be written as
\begin{equation}
P=\frac{1}{4}[1+\tilde{h}(\lambda ,t)]^{2}+\frac{1}{4}\tilde{g}^{2}(\lambda
,t)+\frac{1}{2}\tilde{g}(\lambda ,t),  \label{18}
\end{equation}%
where%
\begin{eqnarray*}
\tilde{g}(\lambda ,t) &=&\tilde{a}^{4}+\tilde{b}^{4}+2\tilde{a}^{2}\tilde{b}%
^{2}\cos [(\tilde{x}_{1}-\tilde{x}_{2})t], \\
\tilde{h}(\lambda ,t) &=&2\tilde{a}^{2}\tilde{b}^{2}\{1-\cos [(\tilde{x}_{1}-%
\tilde{x}_{2})t]\}.
\end{eqnarray*}%
The new parameters $\tilde{x}_{1}$, $\tilde{x}_{2}$, $\tilde{a}$, and $%
\tilde{b}$ are given by $\tilde{F}=F[N/2-1]$ with $\tilde{F}=\tilde{x}_{1}$,
$\tilde{x}_{2}$, $\tilde{a}$, and $\tilde{b}$.\ \

To obtain the exact result of the purity of the central qubit, we need to
know the coupling strength $\lambda ^{\prime }$ \cite{Coupling strength}
between the central qubit and the bath. In different references, this
coupling strength is treated differently. For example, this coupling
strength was assumed to be inversely proportional to the qubit number of the
bath in Ref. \cite{Wezel}, while it was assumed to be inversely proportional
to the square root of the spin number in some other references \cite%
{Paganelli,Square root}. We hereafter denote the two cases with
the above two different coupling strengths as Cases I and II,
respectively. Generally speaking, when the central qubit is
identical to the qubits of the bath, the coupling strength between
the central qubit and the bath should be equal to the coupling
strength between the qubits of the bath. This is Case I, and we
will elaborate it in next subsection. Besides Case I, Case II will
also be discussed later. The calculation for Case II is the same
as that for Case I except that $\lambda ^{\prime }$ in Case I is
changed to $\sqrt{N}\lambda ^{\prime }$. The two different
coupling strengths will lead to different behaviors of the purity.
\subsection{The coupling strength inversely proportional to the spin number
of the bath (Case I)}

Firstly let us consider Case I. Similar to the coupling mechanism
in Ref. \cite{Wezel}, we assume that the coupling strength
$\lambda ^{\prime }$ is just the coupling strength $\lambda /N$
between any two qubits of the bath. Since the coupling strength
$\lambda /N$ between qubits in the LMG model is inversely
proportional to the spin number $N$, the system is extensive.
\begin{figure}[h]
\begin{center}
\includegraphics[bb=90 2 320 161, width=6cm, clip]{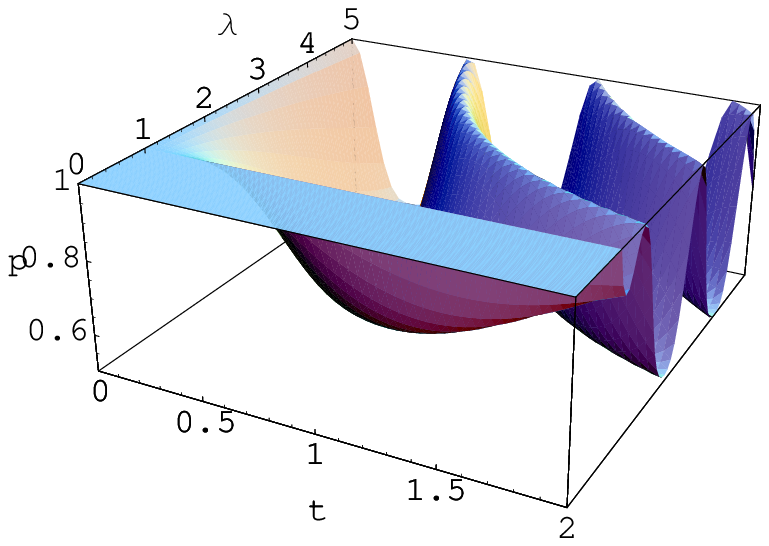} %
\includegraphics[bb=91 1 321 185, width=6cm, clip]{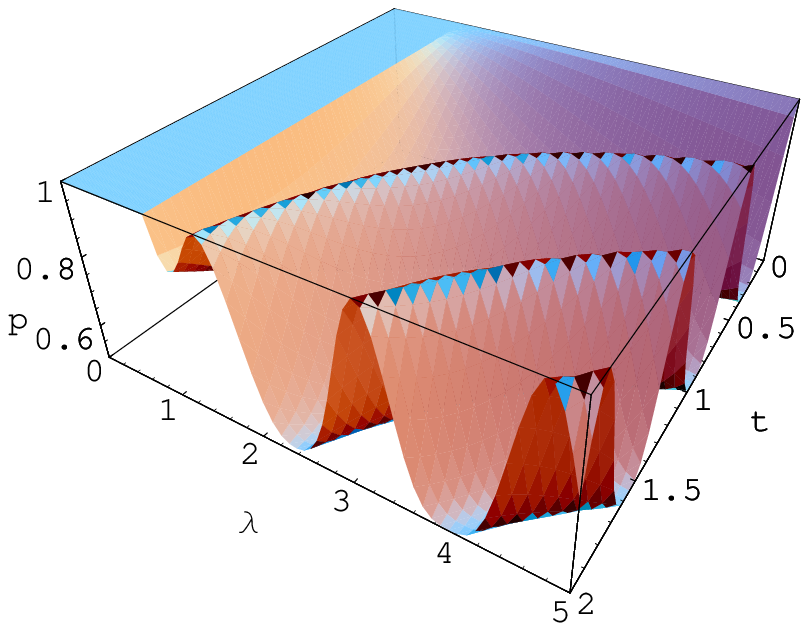}
\end{center}
\caption{(Color online) Two different view angles on the dynamic evolution
of the purity $P$ as functions of $\protect\lambda $ and $t$ (Case I). The
QPT (symmetry breaking) at $\protect\lambda =1$ is well signatured. The
purity saturates when $N$ becomes large. Here we have chosen the qubit
number of the environment $N=5000$.}
\end{figure}

\begin{figure}[h]
\begin{center}
\includegraphics[bb=90 2 321 145, width=8cm, clip]{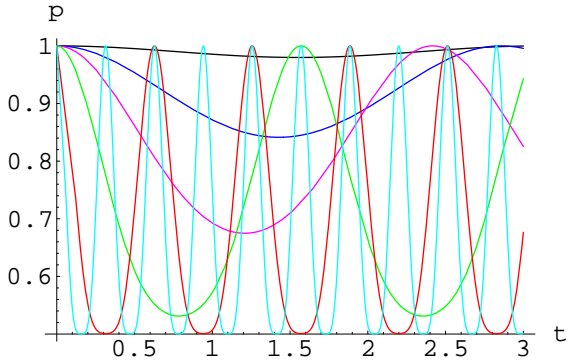}
\end{center}
\caption{(Color online) Dynamic evolution of the purity $P$ as a function of
time $t$ for different $\protect\lambda $. In the symmetry broken phase $(0<%
\protect\lambda <1)$, $P$ remains as a constant unity. In the symmetric
phase $(\protect\lambda >1)$, $P$ ranges from $0.5$ to $1$, and the period
of the oscillation decreases as $\protect\lambda $ increases. The curves
with different colors represent $\protect\lambda =1.01$, $\protect\lambda %
=1.1$, $\protect\lambda =1.3$, $\protect\lambda =2$, and $\protect\lambda =5$
($N=5000$).}
\end{figure}

Fig. 1 clearly shows that, in the symmetry breaking phase, the
purity of the central qubit remains as a constant unity, i.e., the
qubit preserves its quantum coherence all the time. The decay of
purity becomes vanishingly small in the thermodynamic limit. While
in the symmetric phase, the purity varies periodically, as shown
in Fig. 2.

The constant purity $P=1$ of the central qubit for the symmetry broken phase
of the bath can also be verified through another approach. The ground state
of the LMG model in the symmetry broken phase $(0<\lambda <1)$ is $%
\left\vert G\right\rangle =\left\vert N/2,N/2\right\rangle $. The direct
product of the bath $\left\vert G\right\rangle $ and the central qubit $%
c_{\uparrow }\left\vert \uparrow \right\rangle +c_{\downarrow }\left\vert
\downarrow \right\rangle $ can be expanded in the angular momentum coupling
representation%
\begin{eqnarray}
\left\vert G\right\rangle \otimes (c_{\uparrow }\left\vert \uparrow
\right\rangle +c_{2}\left\vert \downarrow \right\rangle ) &\approx
&c_{\uparrow }\left\vert \frac{1}{2}(N+1),\frac{1}{2}(N+1)\right\rangle
\label{19} \\
&&+c_{\downarrow }\left\vert \frac{1}{2}(N-1),\frac{1}{2}(N-1)\right\rangle ,
\notag
\end{eqnarray}%
where we have used the Clebsch-Gordan (C-G) coefficient%
\begin{eqnarray}
\left\vert \frac{N}{2},\frac{N}{2}\right\rangle \otimes \left\vert \uparrow
\right\rangle &=&\left\vert \frac{1}{2}(N+1),\frac{1}{2}(N+1)\right\rangle
\label{20} \\
\left\vert \frac{N}{2},\frac{N}{2}\right\rangle \otimes \left\vert
\downarrow \right\rangle &=&\frac{\sqrt{N}}{\sqrt{N+1}}\left\vert \frac{1}{2}%
(N-1),\frac{1}{2}(N-1)\right\rangle  \notag \\
&&+\frac{1}{\sqrt{N+1}}\left\vert \frac{1}{2}(N+1),\frac{1}{2}%
(N-1)\right\rangle .  \notag
\end{eqnarray}%
The total Hamiltonian (\ref{5}) of the qubit and the bath can be rewritten as%
\begin{equation}
H=-\frac{\lambda }{N}\left[ 2\mathbf{J}_{N+1}^{2}-2(J_{N+1}^{z})^{2}-(N+1)%
\right] -2J_{N+1}^{z}.  \label{21}
\end{equation}%
Through the above approximation (\ref{19}), both $\left\vert
N/2,N/2\right\rangle \otimes \left\vert \uparrow \right\rangle $ and $%
\left\vert N/2,N/2\right\rangle \otimes \left\vert \downarrow \right\rangle $
are the eigenstate of the total Hamiltonian with the eigenenergy $-(N+1)$
and $2\lambda /N-(N-1)$ respectively, and then the dynamic evolution of the
two states are obvious. After a straightforward derivation, the reduced
density matrix of the qubit is expressed as%
\begin{eqnarray}
\rho ^{S}(t) &=&\left\vert c_{\uparrow }\right\vert ^{2}\left\vert \uparrow
\right\rangle \left\langle \uparrow \right\vert +\left\vert c_{\downarrow
}\right\vert ^{2}\left\vert \downarrow \right\rangle \left\langle \downarrow
\right\vert  \label{22} \\
&&+c_{\uparrow }c_{\downarrow }^{\ast }e^{i(\frac{2\lambda }{N}%
+2)t}\left\vert \uparrow \right\rangle \left\langle \downarrow \right\vert
+h.c.,  \notag
\end{eqnarray}%
and the purity of $\rho ^{S}(t)$ remains as unity by applying Eq. (\ref{15}%
). It is thus proven that the qubit preserves its quantum coherence when the
bath is in its symmetry broken phase $(0<\lambda <1)$.

\subsection{The coupling strength inversely proportional to the square root
of the spin number of the bath (Case II)}

We now turn to Case II \cite{Paganelli, Square root}. Being
different from Case I, the purity of the central qubit does not
preserve its coherence when the bath is in the symmetry broken
phase (see Fig.3), although the purity varies also periodically in
the symmetric phase (Fig.4). However, both the range and the
pattern of its time-dependance in the two phases are different
from those in Case I. We also remark that the purity $P$ saturates
in the symmetry broken phase when $N$ increases, just like that in
the symmetric phase of Case I; while the dynamic behavior of the
purity in the symmetric phase depends on $N$ with the period being
inversely proportional to $\sqrt{N}$ approximately, as analyzed
later.

\begin{figure}[h]
\begin{center}
\includegraphics[bb=90 1 321 185, width=6cm, clip]{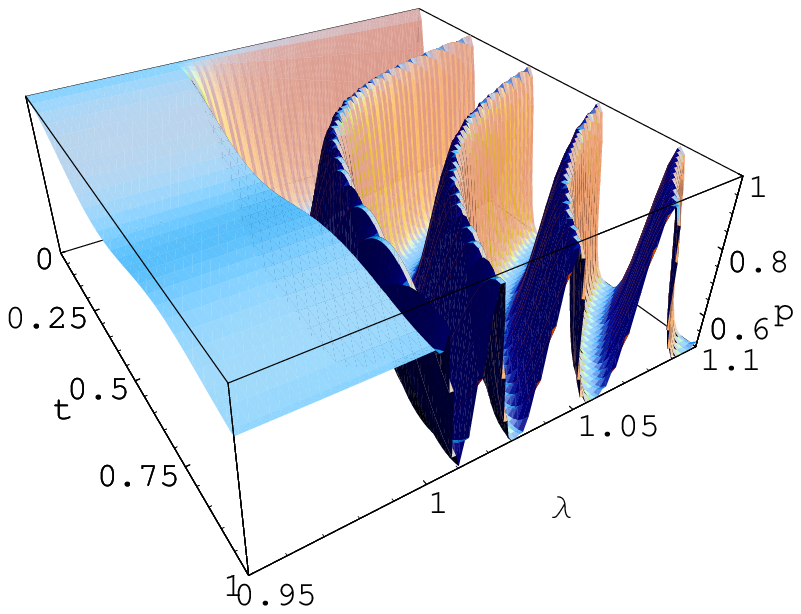} %
\includegraphics[bb=90 1 322 191, width=6cm, clip]{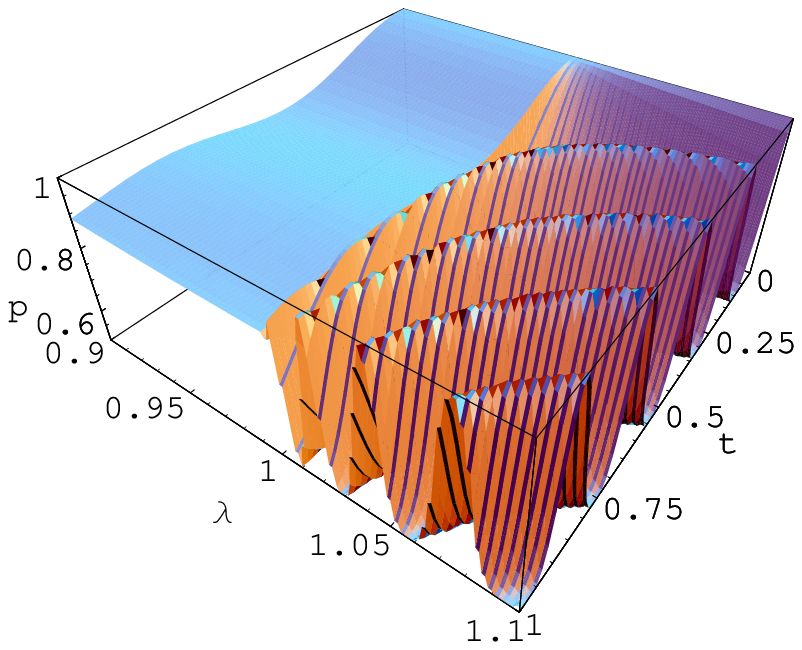}
\end{center}
\caption{(Color online) Dynamic evolution of the purity $P$ as functions of $%
\protect\lambda $ and $t$ (Case II) in the symmetric phase and
symmetry broken phase. Clearly the purity varies in distinctly
different manners in the two phases, which may be considered as an
indication of QPT at the critical point $\protect\lambda =1$. The
purity reaches a steady state in the symmetry broken phase, while
the pattern of the purity will always change with $N$ in the
symmetric phase. Here we choose $N=1000$.}
\end{figure}

\begin{figure}[h]
\begin{center}
\includegraphics[bb=90 1 321 145, width=8cm, clip]{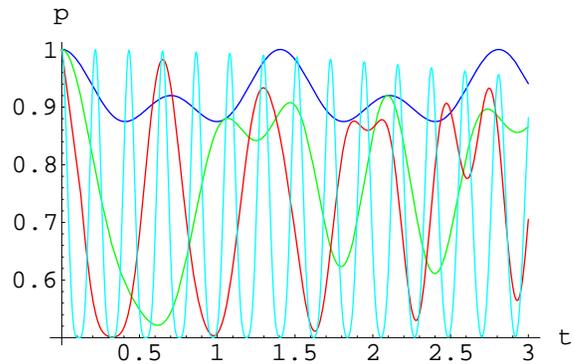}
\end{center}
\caption{(Color online) Dynamic evolution of the purity $P$ as a function of
time $t$ for different $\protect\lambda $. In the symmetry breaking phase $%
(0<\protect\lambda <1)$, the dynamic behavior of $P$ saturates when $N$
increases, which ranges from about $0.88$ to $1$ with the period about $1.4$%
. In the symmetric phase $(\protect\lambda >1)$, $P$ ranges from $0.5$ to $1$%
, and the period of the oscillation decreases as $\protect\lambda $
increases. The curves of different colors represent $\protect\lambda =1.0001$%
, $\protect\lambda =1.0003$, $\protect\lambda =1.002$, $\protect\lambda %
=1.02 $ ($N=1000$).}
\end{figure}

\section{EQUIVALENCE TO AN ANTI JAYNES-CUMMINGS MODEL}

In this section we show the equivalence between the above model
and an anti J-C model with an intensity-dependent coupling
strength \cite{intensity}. With this observation, we can exactly
solve the dynamical equation about time evolution.

\subsection{Symmetry broken phase}

When the bath is in the symmetry broken phase, our model may be recast into
an anti J-C model. Actually the equivalence between the LMG model and Dike
model was just studied recently \cite{Equivalence}.

In the symmetry broken phase $(0<\lambda <1)$, the ground state $\left\vert
N/2,N/2\right\rangle $ of the bath corresponds to a low excitation Fock
state $\left\vert 0\right\rangle $ after the H-P transformation. The mean
photon number $n=\left\langle d^{\dagger }d\right\rangle =0$. Hence we can
directly expand the Holstein-Primakoff (H-P) transformation \cite{H-P} to
the first-order \cite{jin}%
\begin{eqnarray}
J_{N}^{+} &=&\sqrt{N}d,J_{N}^{-}=(J_{N}^{+})^{\dagger }  \notag \\
J_{N}^{z} &=&N/2-d^{\dagger }d,  \notag
\end{eqnarray}%
and the Hamiltonian (\ref{6}) can be rewritten as%
\begin{equation}
H=2(1-\lambda )d^{\dagger }d-N-2\lambda ^{\prime }\sqrt{N}(s_{+}d^{\dagger
}+s_{-}d)-2s_{z}.  \label{24}
\end{equation}%
Let us recall that the anti J-C Hamiltonian can be rewritten as%
\begin{equation}
H_{AJC}=\nu d^{\dagger }d-k(\sigma _{+}d^{\dagger }+\sigma _{-}d)+\frac{1}{2}%
\omega \sigma _{z},  \label{23}
\end{equation}%
where $d^{\dagger }$ and $d$ are the creation and annihilation operators of
the single-mode quantized field with the frequency $\nu $; $-k$ is the
coupling strength between the field and the two-level system; $\omega $ is
the level spacing between the two-level system; $\sigma _{+}=(\sigma
_{x}+i\sigma _{y})/2$ and $\sigma _{-}=(\sigma _{x}-i\sigma _{y})/2$. Hence,
the model described by Eq. (\ref{24}) is an anti J-C model (\ref{23}) with $%
\nu =2(1-\lambda )$, $\omega =-2$, and $k=2\lambda ^{\prime }\sqrt{N}$. We
now illustrate that the boson mode characterized by $d$ and $d^{\dagger }$
may be mapped from the collective spin $J$ in a low excitation limit. Note
that the different mapping ways depend on the phases of the bath because the
bosonization of collective spin is essentially a mean field approach based
on choice of the order parameter.

The solution $\left\vert \Psi (t)\right\rangle $ of the Schrodinger equation
$i\hbar \partial _{t}\left\vert \Psi (t)\right\rangle =H_{AJC}\left\vert
\Psi (t)\right\rangle $ can be expressed as%
\begin{equation}
\left\vert \Psi (t)\right\rangle =\sum_{n=0}[c_{\uparrow ,n+1}(t)\left\vert
\uparrow \right\rangle \otimes \left\vert n+1\right\rangle +c_{\downarrow
,n}(t)\left\vert \downarrow \right\rangle \otimes \left\vert n\right\rangle
],  \label{23.1}
\end{equation}%
where $n=\left\langle d^{\dagger }d\right\rangle $ is the mean
\textquotedblleft photon" number. A straightforward calculation determines
the probability amplitudes \cite{J-C model}%
\begin{eqnarray}
c_{\uparrow ,n+1}(t) &=&\{c_{\uparrow ,n+1}(0)\left[ \cos \left( \Omega
_{n}t\right) -i\frac{\Delta }{\Omega _{n}}\sin (\Omega _{n}t)\right]
\label{23.2} \\
&&+i\frac{k}{\Omega _{n}}\sqrt{n+1}c_{\downarrow ,n}(0)\sin \left( \Omega
_{n}t\right) \}\exp \left( i\frac{\Delta }{2}t\right) ,  \notag
\end{eqnarray}%
\begin{eqnarray}
c_{\downarrow ,n}(t) &=&\{c_{\downarrow ,n}(0)\left[ \cos \left( \Omega
_{n}t\right) +i\frac{\Delta }{\Omega _{n}}\sin \left( \Omega _{n}t\right) %
\right]  \label{23.3} \\
&&+i\frac{k}{\Omega _{n}}\sqrt{n+1}c_{\uparrow ,n+1}(0)\sin \left( \Omega
_{n}t\right) \}\exp \left( -i\frac{\Delta }{2}t\right) ,  \notag
\end{eqnarray}%
where $\Delta =\nu +\omega $ and $\Omega _{n}=\sqrt{(\Delta
/2)^{2}+k^{2}(n+1)}$.

After a routine calculation, we obtain the purity of the central qubit (with
an initial state $(\left\vert \uparrow \right\rangle +\left\vert \downarrow
\right\rangle )/\sqrt{2}$) in the symmetry broken phase $(0<\lambda <1)$
(see Appendix A)
\begin{eqnarray}
P &=&\frac{1}{4}\left[ 1+\cos ^{2}\left( \Omega _{0}t\right) +\left( \frac{%
\Delta }{2\Omega _{0}}\right) ^{2}\sin ^{2}\left( \Omega _{0}t\right) \right]
^{2}  \label{25} \\
&&+\frac{1}{4}\left[ \left( \frac{k}{\Omega _{0}}\right) ^{4}\sin ^{4}\left(
\Omega _{0}t\right) +2\left( \frac{k}{\Omega _{0}}\right) ^{2}\sin
^{2}\left( \Omega _{0}t\right) \right] .  \notag
\end{eqnarray}

In Case I ($\lambda ^{\prime }=\lambda /N$), the coupling strength $%
k=2\lambda /\sqrt{N}$ is inversely proportional to the square root of the
qubit number $N$. In the large $N$ limit,
\begin{equation}
k^{2}=\frac{4\lambda ^{2}}{N}\ll 4\lambda ^{2}=\Delta ^{2},  \label{25.6}
\end{equation}%
i.e.,
\begin{eqnarray}
\lim_{N\rightarrow \infty }\left( \frac{k}{\Omega _{0}}\right) ^{2} &=&\frac{%
4k^{2}}{\Delta ^{2}}=0,  \label{25.7} \\
\lim_{N\rightarrow \infty }\left( \frac{\Delta }{2\Omega _{0}}\right) ^{2}
&=&\frac{\Delta ^{2}}{\Delta ^{2}}=1.  \notag
\end{eqnarray}%
Hence from Eq. (\ref{25}), we have $P=1$ in large $N$ limit. This analytical
analysis agrees well with Eq. (\ref{22}) and Fig. 1.

Generally speaking, the quantum coherence (measured by purity) of a quantum
open system would be dissipated by its bath. But when the coupling strength
between the system and its bath becomes vanishingly small, the system and
bath becomes decoupled, and then the system will preserve all its coherence
(remains in a pure state or the purity remains to be unity) during the
dynamic evolution. In the thermodynamic limit, the coupling strength between
the two level atom and the \textquotedblleft radiation field" becomes
vanishingly small, i.e., the radiation field and the central qubit are
decoupled. Thus, the central qubit evolves under the free Hamiltonian $%
H_{S}=-2s_{z}$, which preserves quantum coherence of the central qubit.

In Case II ($\lambda ^{\prime }=\lambda /\sqrt{N}$), the coupling strength $%
k=2\lambda $.\ Even in the thermodynamic limit, the interaction Hamiltonian
does not vanish ($2\lambda ^{\prime }\sqrt{N}=2\lambda \neq 0$). This is why
the purity of the central qubit does not preserve in Case II even when the
bath is in the symmetry broken phase. In this case, the purity (\ref{25})
can be simplified to%
\begin{equation}
\lim_{N\rightarrow \infty }P=\frac{1}{4}\left[ \frac{32}{25}\sin ^{4}(\sqrt{5%
}\lambda t)-\frac{8}{5}\sin ^{2}(\sqrt{5}\lambda t)+4\right] ,  \label{26.1}
\end{equation}%
which is independent of $N$, as we observed numerically in Sec. IV.

\subsection{Symmetric phase}

In the above discussion, the system and the bath is reduced into the anti
J-C model with a normal coupling, i.e., the coupling does not depend on the
mean \textquotedblleft photon" number. We now show that the system and the
bath can be reduced into the anti J-C model with an intensity-dependent
coupling when the bath is in the symmetric phase. In the symmetric phase $%
(\lambda >1) $, however, the ground state $\left\vert N/2,I\left( \lambda
\right) \right\rangle $ of the bath is no longer a low excitation state
after the H-P transformation. The mean \textquotedblleft photon" number%
\begin{equation}
n=\left\langle d^{\dagger }d\right\rangle =\frac{N}{2}-I(\lambda )\approx
\frac{N}{2}\left( 1-\frac{1}{\lambda }\right) .  \label{27}
\end{equation}%
is of the same order of $N$. By applying H-P transformation the Hamiltonian (%
\ref{6}) can be rewritten as
\begin{eqnarray}
H &=&\frac{2\lambda }{N}(d^{\dagger }d)^{2}+2(1-\lambda )d^{\dagger }d-N
\label{29} \\
&&-2\lambda ^{\prime }(s_{+}d^{\dagger }\sqrt{N-d^{\dagger }d}+s_{-}\sqrt{%
N-d^{\dagger }d}d)-2s_{z}.  \notag
\end{eqnarray}%
The model described by this Hamiltonian (\ref{29}) is an intensity-dependent
coupling anti J-C model with a Kerr-effect term $2\lambda (d^{\dagger
}d)^{2}/N$, which can be analytically diagonalized as well. After a similar
derivation to that in the symmetry broken phase, we obtain the solution $%
\left\vert \Phi (t)\right\rangle $ of the Schrodinger equation $i\hbar
\partial _{t}\left\vert \Phi (t)\right\rangle =H\left\vert \Phi
(t)\right\rangle $,%
\begin{equation}
\left\vert \Phi (t)\right\rangle =\sum_{n=0}[c_{\uparrow ,n+1}^{\prime
}(t)\left\vert \uparrow \right\rangle \otimes \left\vert n+1\right\rangle
+c_{\downarrow ,n}^{\prime }(t)\left\vert \downarrow \right\rangle \otimes
\left\vert n\right\rangle ],  \label{30}
\end{equation}%
where

\begin{eqnarray}
&&c_{\uparrow ,n+1}^{\prime }(t) \\
&=&\{c_{\uparrow ,n+1}^{\prime }(0)\left[ \cos \left( \Omega _{n}^{\prime
}t\right) -i\frac{\Lambda _{n}^{\prime }}{\Omega _{n}^{\prime }}\sin \left(
\Omega _{n}^{\prime }t\right) \right]   \notag \\
&&+ic_{\downarrow ,n}^{\prime }(0)\frac{\Gamma _{n}^{\prime }}{\Omega
_{n}^{\prime }}\sin \left( \Omega _{n}^{\prime }t\right) \}\exp \left\{
-iA_{n}t\right\}   \notag \\
&&c_{\downarrow ,n}^{\prime }(t)  \notag \\
&=&\{c_{\downarrow ,n}^{\prime }(0)\left[ \cos \left( \Omega _{n}^{\prime
}t\right) +i\frac{\Lambda _{n}^{\prime }}{\Omega _{n}^{\prime }}\sin \left(
\Omega _{n}^{\prime }t\right) \right]   \notag \\
&&+ic_{\uparrow ,n+1}^{\prime }(0)\frac{\Gamma _{n}^{\prime }}{\Omega
_{n}^{\prime }}\sin \left( \Omega _{n}^{\prime }t\right) \}\exp \left\{
-iB_{n}t\right\} ,  \notag
\end{eqnarray}%
and%
\begin{eqnarray}
\Delta ^{\prime } &=&2(1-\lambda )+(-2)=-2\lambda , \\
\Lambda _{n}^{\prime } &=&\frac{\lambda }{N}(2n+1)+\frac{\Delta ^{\prime }}{2%
},  \notag \\
\Gamma _{n}^{\prime } &=&2\lambda ^{\prime }\sqrt{(N-n)(n+1)},  \notag \\
\Omega _{n}^{\prime } &=&\sqrt{(\Lambda ^{\prime })^{2}+(\Gamma ^{\prime
})^{2}},  \notag \\
A_{n} &=&\frac{\lambda }{N}\left[ n^{2}+(n+1)^{2}\right] -\frac{\Delta
^{\prime }}{2},  \notag \\
B_{n} &=&\frac{\lambda }{N}\left[ n^{2}+(n+1)^{2}\right] +\frac{\Delta
^{\prime }}{2}.  \notag
\end{eqnarray}%
The purity of the central qubit (with an initial state $(\left\vert \uparrow
\right\rangle +\left\vert \downarrow \right\rangle )/\sqrt{2}$) in the
symmetric phase $(\lambda >1)$ can be determined as (see Appendix B)%
\begin{equation}
P=\frac{1}{2}+\frac{1}{2}\left[ 1-\left( \frac{2\lambda ^{\prime }}{\Omega
_{n}^{\prime }}\right) ^{2}(N-n)(n+1)\sin ^{2}\left( \Omega _{n}^{\prime
}\right) \right] ^{2}.  \label{32}
\end{equation}

In Case I ($\lambda ^{\prime }=\lambda /N$), the purity $P$ (\ref{32}) can
be further simplified as%
\begin{equation}
P=\frac{1}{2}+\frac{1}{2}\left[ 1-\left( 1-\frac{1}{\lambda ^{2}}\right)
\sin ^{2}\left( \lambda t\right) \right] ^{2}.  \label{33}
\end{equation}
We see from Eq. (\ref{33}) that $P$ varies periodically and is independent
of $N$ though the coupling strength $-2\lambda ^{\prime }\sqrt{N-d^{\dagger
}d}$ in Eq. (\ref{30}), as we have seen in Fig. 1. The physics behind Eq. (%
\ref{33}) is that the mean \textquotedblleft photon" number of the ground
state is also $N$-dependent, which countervails with the $N$-dependent
coupling strength, leading to the $N$-independent dynamical behavior of the
purity $P$ in Case I.

On the other hand, in Case II ($\lambda ^{\prime }=\lambda /\sqrt{N}$), the
purity $P$ (\ref{32}) can be further simplified as%
\begin{equation}
P=\frac{1}{2}+\frac{1}{2}\left\{ 1-\sin ^{2}\left[ \sqrt{N(\lambda ^{2}-1)}t%
\right] \right\} ^{2}.  \label{34}
\end{equation}%
Hence, the behavior of purity would not reach a steady pattern when $N$
increases, as observed in Figs. 3 and 4.\ The $N$-dependence of the purity $%
P $ in Case II\ stens from the coupling strength $-2\lambda ^{\prime }\sqrt{%
N-d^{\dagger }d}$ and the $N$-dependent mean \textquotedblleft
photon" number of the ground state; the $N$-dependence of them
cannot countervail with each other.

\section{SUMMARY}

We have studied the dynamic property of a central qubit coupled to an
isotropic Lipkin-Meshkov-Glick bath. Two different types of coupling
strength between the central qubit and the bath are considered. In both
cases, the QPT of the bath is well revealed by the dynamic behavior of the
central qubit. We have found that our model is equivalent to an anti J-C
model under H-P transformation when the bath is in the symmetry broken
phase. Especially, when the coupling strength between the central qubit and
the bath is inversely proportional to the spin number of the bath, the
central spin and the bath becomes decoupled, and the central qubit preserves
its quantum coherence all the time. The present study not only demonstrates
how the QPT influence the quantum coherence of the central qubit, but also
establishes the connection between the LMG model and anti J-C model. In
addition, our investigation may propose a new scenario to preserve quantum
coherence of a central qubit in experimental implementation of quantum
computation.

\textbf{Acknowledgement:} This work was supported by the RGC grants of Hong
Kong (HKU-3/05C and HKU 7051/06P), and Seed Funding grants of HKU, the
National Natural Science Foundation of China under Nos. 10429401, 90203018,
10474104, and 60433050, and the state key programs of China under Nos.
2001CB309310, 2005CB724508, and 2006CB0L1001.

\begin{appendix}

\setcounter{section}{0} \setcounter{equation}{0} \renewcommand{\thesection}{%
\Alph{section}}
\appendix

\section {DYNAMICS OF PURITY IN THE SYMMETRY BROKEN PHASE ($0<\lambda <1 $)}

For the anti J-C Hamiltonian (\ref{23}), the time evolution of a
initial state $\left\vert \Psi (0)\right\rangle =(\left\vert
\uparrow \right\rangle +\left\vert \downarrow \right\rangle
)/\sqrt{2}\otimes \left\vert
0\right\rangle $ can be expressed as%
\begin{eqnarray}
\left\vert \Psi (t)\right\rangle  &=&\frac{ik}{\sqrt{2}\Omega
_{0}}\sin \left( \Omega _{0}t\right) \}\exp \left( i\frac{\Delta
}{2}t\right) \left\vert \uparrow \right\rangle \otimes \left\vert
1\right\rangle
\label{35} \\
&&+\left[ \cos \left( \Omega _{0}t\right) +\frac{i\Delta }{2\Omega
_{0}}\sin
\left( \Omega _{0}t\right) \right]   \notag \\
&&\times \exp \left( -\frac{i\Delta }{2}t\right) \left\vert
\downarrow \right\rangle \otimes \left\vert 0\right\rangle
+\frac{\left\vert \uparrow \right\rangle }{\sqrt{2}}\otimes
\left\vert 0\right\rangle .  \notag
\end{eqnarray}%
The reduced density matrix $\rho ^{S}(t)$ (Eq.(\ref{14}))\ of the
system is then
found to be%
\begin{eqnarray}
&&\rho ^{S}(t)  \label{36} \\
&=&\mathrm{Tr_{B}}\left\vert \Psi (t)\right\rangle \left\langle
\Psi
(t)\right\vert   \notag \\
&=&\frac{1}{2}\left[ 1+\left( \frac{k}{\Omega _{0}}\right)
^{2}\sin ^{2}\left( \Omega _{0}t\right) \right] \left\vert
\uparrow \right\rangle
\left\langle \uparrow \right\vert   \notag \\
&&+\frac{1}{2}\left[ \cos ^{2}\left( \Omega _{0}t\right) +\left( \frac{%
\Delta }{2\Omega _{0}}\right) ^{2}\sin ^{2}\left( \Omega
_{0}t\right) \right] \left\vert \downarrow \right\rangle
\left\langle \downarrow \right\vert
\notag \\
&&+\frac{1}{2}\left[ \cos \left( \Omega _{0}t\right) -i\frac{\Delta }{%
2\Omega _{0}}\sin \left( \Omega _{0}t\right) \right] \exp \left( i\frac{%
\Delta }{2}t\right) \left\vert \uparrow \right\rangle \left\langle
\downarrow \right\vert   \notag \\
&&+h.c.  \notag
\end{eqnarray}%
Applying Eq. (\ref{15}), we obtain the purity $P$ (Eq.(\ref{25})) of
the
central qubit%
\begin{eqnarray}
P &=&\frac{1}{4}\left[ 1+\cos ^{2}\left( \Omega _{0}t\right) +\left( \frac{%
\Delta }{2\Omega _{0}}\right) ^{2}\sin ^{2}\left( \Omega
_{0}t\right) \right]
^{2}  \label{37} \\
&&+\frac{1}{4}\left[ \left( \frac{k}{\Omega _{0}}\right) ^{4}\sin
^{4}\left( \Omega _{0}t\right) +2\left( \frac{k}{\Omega
_{0}}\right) ^{2}\sin ^{2}\left( \Omega _{0}t\right) \right] .
\notag
\end{eqnarray}

\section {DYNAMICS OF PURITY IN THE SYMMETRIC PHASE ($\lambda >1$)}

For a generalized anti J-C Hamiltonian (Eq.(\ref{29})), the time
evolution of an initial state $\left\vert \Phi (0)\right\rangle
=(\left\vert \uparrow \right\rangle +\left\vert \downarrow
\right\rangle )/\sqrt{2}\otimes \left\vert n\right\rangle $ can be
expressed as

\begin{eqnarray}
\left\vert \Phi (t)\right\rangle  &=&\frac{1}{\sqrt{2}}\left\{
\cos \left(
\Omega _{n-1}^{\prime }t\right) -i\frac{\Xi _{n-1}}{\Omega _{n-1}^{\prime }}%
\sin \left( \Omega _{n-1}^{\prime }t\right) \right\}   \label{38} \\
&&\exp \left( -iA_{n-1}t\right) \left\vert \uparrow \right\rangle
\otimes
\left\vert n\right\rangle +\frac{i}{\sqrt{2}}\frac{\Gamma _{n}^{\prime }}{%
\Omega _{n-1}^{\prime }}\sin \left( \Omega _{n-1}^{\prime
}t\right)   \notag
\\
&&\exp \left( -iB_{n-1}t\right) \left\vert \downarrow
\right\rangle \otimes
\left\vert n-1\right\rangle   \notag \\
&&+\frac{1}{\sqrt{2}}\left\{ \cos \left( \Omega _{n-1}^{\prime }t\right) +i%
\frac{\Xi _{n}}{\Omega _{n}^{\prime }}\sin \left( \Omega
_{n-1}^{\prime
}t\right) \right\}   \notag \\
&&\exp \left( -iB_{n}t\right) \left\vert \downarrow \right\rangle
\otimes
\left\vert n\right\rangle +\frac{i}{\sqrt{2}}\frac{\Gamma _{n}^{\prime }}{%
\Omega _{n}^{\prime }}\sin \left( \Omega _{n-1}^{\prime }t\right)   \notag \\
&&\exp \left( -iA_{n}t\right) \left\vert \uparrow \right\rangle
\otimes \left\vert n+1\right\rangle ,  \notag
\end{eqnarray}%
where%
\begin{equation*}
\Xi _{n}=\frac{\lambda }{N}(2n+1)+\frac{\Delta ^{\prime }}{2}.
\end{equation*}

In the large $N$ limit, the reduced density matrix $\rho ^{S}(t)$
(Eq.(\ref{14})) of the central qubit is derived as

\begin{eqnarray}
&&\rho ^{S}(t)  \label{39} \\
&=&\mathrm{Tr_{B}}\left\vert \Phi (t)\right\rangle \left\langle
\Phi
(t)\right\vert  \notag \\
&=&\frac{1}{2}\left\vert \uparrow \right\rangle \left\langle
\uparrow \right\vert +\frac{1}{2}\left\vert \downarrow
\right\rangle \left\langle
\downarrow \right\vert +\frac{1}{2}\exp [-i(A_{n}-B_{n})t]\times  \notag \\
&&\left\{ \cos \left( \Omega _{n-1}^{\prime }t\right)
-\frac{i}{\Omega
_{n}^{\prime }}\left[ \frac{\lambda }{N}(2n+1)+\frac{\Delta ^{\prime }}{2}%
\right] \sin \left( \Omega _{n-1}^{\prime }t\right) \right\} ^{2}  \notag \\
&&\times \left\vert \uparrow \right\rangle \left\langle \downarrow
\right\vert +h.c.  \notag
\end{eqnarray}

Using Eq. (\ref{15}), we obtain the purity $P$ (Eq.(\ref{32})) of
the central qubit
\begin{equation}
P=\frac{1}{2}+\frac{1}{2}\left[ 1-\left( \frac{2\lambda ^{\prime
}}{\Omega _{n}^{\prime }}\right) ^{2}(N-n)(n+1)\sin ^{2}\left(
\frac{\Omega _{n}^{\prime }}{2}t\right) \right] ^{2}.  \label{40}
\end{equation}

\end{appendix}

\end{document}